# A probable vacuum state containing a large number of hydrogen atom of excited state or ground state K, Rb or Cs atom

Pei-Lin You

Institute of Quantum Electronics, Guangdong Ocean University, zhanjiang 524025, China.

**The linear Stark effect shows that the first excited state of hydrogen atom has large permanent electric dipole moment (EDM), $d_H = 3ea_o = 1.59 \times 10^{-8}$ e.cm ($a_o$ is Bohr radius). Using special capacitors our experiments discovered that the ground state K, Rb or Cs atom is polar atom with a large EDM of the order of $ea_o$ as hydrogen atom of excited state. Their capacitance(C) at different voltage (V) was measured. The C-V curve shows that the saturation polarization of K, Rb or Cs vapor has be observed when the field $E \geq 10^5$ V/m. When the saturation polarization appeared, nearly all K, Rb or Cs atoms(more than 98 percent) turned toward the direction of the field, and $C \approx C_o$ ($C_o$ is vacuum capacitance) or their dielectric constant is nearly the same as vacuum! K, Rb or Cs vapor just exist in the lowest energy state, so we see the vacuum state containing a large number of atoms! Due to the saturation polarization of hydrogen vapor of excited state is easily appears, we conjecture that a vacuum state containing a large number of hydrogen atom of excited state may exist at early Universe when the temperature T is more than $7.9 \times 10^4$ K.**

**1. Introduction**     New information from upcoming experiments, such as those at the Large Hadron Collider (LHC) at CERN, should help us find answers to many outstanding questions haunting the field of particle physics for few decades. However, what question puzzling us still remain unsolved ? Perhaps, the vacuum state and its details is such a question.

In non-relativistic quantum theory, the ground state was an extremely simple state. It was just the vacuum, the empty world, nothing else, and had therefore the highest possible symmetry [1].

In relativistic quantum theory, Dirac assumed that the normal state of the vacuum then consists of an infinite density of negative-energy electrons. W. Heisenberg once stated that Dirac's theory of the electron has changed the whole picture of atomic physics. In Dirac's theory the ground state was different. It was an object which was filled with particles of negative energy that could not be seen. Besides that, if the process of pair production is introduced one should expect that the ground state must contain probably an infinite number of virtual pairs of positrons and electrons or of particles and antiparticles; so you see at once that the ground state is a complicated dynamical system. It is one of the eigensolutions defined by the underlying natural law and it need not be symmetrical [1].

In particle physics and quantum field theory, vacuum as the source of asymmetry, it is defined as the lowest energy state of a system [2]. F. Wilczek once stated that modern physicists hypothesize that what we perceive as vacuum is actually a highly structured medium. To account for the masses of $W^{\pm}$ and Z, we must suppose that what we perceive as vacuum is, in reality, a new form of superconductor, not for electromagnetism, but for its near-relation gauge interactions. With this interpretation, what we perceive as empty space is not so empty. But what is it that has the role of the Cooper pair whose alignment is responsible for superconductivity? No form of matter identified so far provides a suitable candidate[3].

In Universe physics, it has been suggested that it may indeed be helpful if there exists no parameter-free fundamental physics theory, but rather a family of theories with varying effective parameters(perhaps because of a vast array of possible vacuum states) [4]. P. Coles once stated that in different models of early Universe physics, the low-energy Universe is described by an effective quantum field theory with the configuration of the vacuum state. In these theories there is a myriad of possible vacuum, each with its own set of fundamental constants, different gauge hierarchies, and so on. To produce inflation, a possible way of doing this is by allowing the Universe to undergo a phase transition in which the field rolls down the potential into a new vacuum state. The Universe can be made to accelerate when the field remains in transit between these vacuum states[5]

In the past few decades Dirac's novel idea served as a nucleus for the crystallization of many unrelated ideas and initiated a strong development of theory of the vacuum. Regrettably, none of these speculative theories has received any empirical support from experiment.

In the following few sections we will discuss the orientation polarization of polar molecules (for instance, $H_2O$ molecules) and polar atoms (for instance, Cs atoms). We shall see the difference and relation between the two substances. Finally, we discuss the relation between polar atoms and the vacuum state.

**2. Orientation polarization of polar molecules**   Many molecules do have permanent electric dipole moments(EDM). This molecule is called a polar molecule. A polar molecule tends to be oriented parallel to the filed because of the torque it experiences. The molecular dipole moments are of the order of magnitude of the electronic charge ($1.6 \times 10^{-19}$ C) multiplied by the molecular dimensions ($10^{-10}$ m), or about $10^{-30}$ coulomb-meter.

We considered the orientation polarization of polar molecules in a gas by using statistical mechanics. The local field acting on a molecule in a gas is almost the same as the external field **E**[6]. All molecules are assumed to possess a EDM $\mathbf{d_o}$. In the absence of a field thermal agitation keeps the molecules randomly oriented so that there is no net dipole moment. After switching on a field **E**, a molecule with dipole moment acquires an additional potential energy $U = -d_o E \cos\theta$, where $\theta$ is the angle between $\mathbf{d_o}$ and **E**. The probability distribution at thermal equilibrium is now modified by the presence of a Boltzmann factor $\exp(-U/kT)$, consequently there will be an average dipole moment. In addition, all the molecules also acquire an induced dipole moment. If the molecular polarizability is $\alpha$, there is an additional induced polarization $N\alpha\varepsilon_o E$, where N is the number density of molecules. Finally, we have the result that the polarization, the average dipole moment per unit volume, is

$$P = N \alpha \varepsilon_o E + N d_o L(a) \tag{1}$$

The electric susceptibility of a gaseous polar dielectric is[7]

$$x_e = N \alpha + N d_o L(a) / \varepsilon_o E \tag{2}$$

where $a = d_o E /kT$, k is Boltzmann constant, $\varepsilon_o$ is the permittivity of free space. the Langevin function is

$$L(a) = [(e^a + e^{-a}) / (e^a - e^{-a})] - 1/a \tag{3}$$

The Langevin function L(a) is equal to the mean value of $\cos\theta$ [7]:

$$<\cos\theta> = \mu \int_0^\pi \cos\theta \exp(d_o E \cos\theta /kT) \sin\theta \, d\theta = L(a) \tag{4}$$

where $\mu$ is a normalized constant. This result shows that L(a) is the percentage of polar molecule lined up with the field in the total number. When $a \ll 1$ and $L(a) \approx a/3$, when $a \gg 1$ and $L(a) \approx 1$. For all polar molecules $d_o E/kT$ is normally very small, so $L(a) \approx a/3 = d_o E/3kT$, we have

$$x_e = N \alpha + N d_o^2 /3k \varepsilon_o T = A + B/T. \tag{5}$$

where A and the slope B is constant at a fixed density. Note that the electric susceptibility caused by the orientation of polar molecules is inversely proportional to the absolute temperature: $x_e = B/T$ while the induced electric susceptibility due to the distortion of electronic motion in atoms or molecules is temperature independent: $x_e = A$. Clearly, this difference in temperature dependence offers a means of separating the polar and non-polar substances experimentally. Because of $x_e = C/C_o - 1$, where $C_o$ is the vacuum capacitance and C is the capacitance of the capacitor filled with the material, by measuring the capacitance C at different temperature T, it is possible to distinguish between permanent and induced dipole moment.

The electric susceptibility caused by the orientation of gaseous polar molecules is given by the expression

$$x_e = N d_o^2 /3kT \varepsilon_o \tag{6}$$

The slope is $B = N d_o^2 /3k \varepsilon_o$ and the EDM of a molecule is

$$d_o = (3k \varepsilon_o B / N)^{1/2} \tag{7}$$

R.P. Feynman considered the orientation polarization of water vapor molecules [8]. He plotted the straight line from four experimental points is

$$x_e = A + B/T \approx 0.002 + B/T \tag{8}$$

Table 1 gives the four experimental data of water vapor at various temperatures [9]. From the ideal gas law, the average density of water vapor is $N = P/kT = (1.392 \pm 0.002) \times 10^{25}$ m$^{-3}$. We work out the slope of the line B = $(1.50 \pm 0.04)$K and the EDM of a water molecule

$$d_{H2O}= (3k \varepsilon_o B/N)^{1/2} = (6.28\pm0.18)\times 10^{-30} \text{ C.m} \quad (9)$$

**Table 1 Experimental measurements of the susceptibility of water vapor at various temperatures**

| T(K) | Pressure(cm Hg) | $x_e$ |
|---|---|---|
| 393 | 56.49 | 0.004002 |
| 423 | 60.93 | 0.003717 |
| 453 | 65.34 | 0.003488 |
| 483 | 69.75 | 0.003287 |

The result in agreement with the observed value $d_{H2O}= 6.20\times 10^{-30}$ C. m[6].

### 3. Orientation polarization of polar atoms

**A. Theoretical motivation** In order for an atom or elementary particle to possess a EDM, time reversal (T) symmetry must be violated, and through the CPT theorem CP(charge conjugation and parity) must be violated as well[10]. Search for an EDM of atom, the most sensitive of which is done with $^{199}$Hg(the result is d(Hg)=-[1.06±0.49 (stat)±0.40 (syst)]×$10^{-28}$e.cm )[10]. Experiments to search for an EDM of atom for more than 50 years, no large EDM has been obtained so far ($d_{atom} \ll 10^{-24}$ e.cm) [10-12]. The linear Stark effect shows that the first excited state of hydrogen atom has very large EDM, $d_H$=3e $a_o$=1.59×$10^{-8}$e.cm ($a_o$ =0.53×$10^{-8}$ cm is Bohr radius) [13,14]. The radius of the hydrogen atom is $r_H$= 4$a_o$ =2.12×$10^{-8}$ cm, it is almost the same as the $^{199}$Hg ($r_{Hg}$ =1.51×$10^{-8}$ cm) [15], but the discrepancy between their EDM is by some twenty orders of magnitude! How do explain this inconceivable discrepancy? The existing theory can not answer the problem! The existing theory thinks that in quantum mechanics there is no such concept as the path of an electron[14]. No one will give you any deeper explain of the inconceivable discrepancy! However, a hydrogen atom (n≧2) has a nonzero EDM in the semi-classical theory of atom. The electron in a hydrogen atom (n≧2) moves along a quantization elliptic orbit. We can draw a straight line perpendicular to the major axis of the elliptic orbit through the nucleus in the orbital plane. The straight line divides the elliptic orbit into two parts that are different in size. According conservation of the area velocity, the average distance between the moving electron and the static nucleus is larger and the electron remains in the large part longer than in the small part. As a result, the time-averaged value of the electric dipole moment over a period is nonzero.

In past experiments, they measured microcosmic Larmor precession frequency of individual atom based on weak interactions [10-12]. We have submitted three articles on the similar topic, however, with measuring macroscopic electric susceptibility($\chi_e$) of K, Rb or Cs vapor containing a large number of atoms based on electromagnetic interactions. Our experiments showed that a ground state neutral K, Rb or Cs atom is polar atom with a large EDM of the order of e $a_o$ as excited state of hydrogen atom[16-18]. These results are the product of eight years of intense research [19, 20]. Few experiments in atomic physics have produced a result as surprising as this one!

**B. Experimental method and result of Cs atom** The first experiment : investigation of the relationship between the capacitance C' of Cs or Hg vapor and their density N. The experimental apparatus were two closed glass containers containing Cs or Hg vapor. Two silver layers **a** and **b** build up the glass cylindrical capacitor(Fig.1,where the plate area $S_1$=(5.8±0.1)×$10^{-2}$ m$^2$, the plate separation $H_1$=(9.6± 0.1)mm). The radiuses of the layers **a** and **b** are shown respectively by r and R. Since R-r=H+2△<< r (△ =1.2mm), the capacitor can be approximately regarded as a parallel-plane capacitor. The capacitance was measured by a digital meter. The precision of the meter was 0.1pF, the accuracy was 0.5% and the surveying voltage was V=(1.0±0.01)volt. It means that the applied field E=V/$H_1$≈1.1×$10^2$v/m is weak using the meter. This capacitor is equivalently connected in series by two capacitors. One is called C' and contains the Cs or Hg vapor of thickness H , another one is called C'' and contains the glass medium of thickness 2△. The total capacitance C can be written as C=C'C''/ (C'+C'') or C'=CC''/ (C''−C), where C'' and C can be directly measured. When the two containers are empty, they are pumped to a vacuum pressure P≦$10^{-6}$ Pa for 20 hours. The measured capacitances are nearly the same: C'$_{10}$=(54.0±0.1)pF(for Cs) and C'$_{20}$=(52.8±0.1)pF(for Hg). Next, a small amount of the Cs or Hg material with high purity is put in the two containers respectively in a vacuum environment. The two containers are again pumped to vacuum pressure P ≤10$^{-6}$ Pa , then they are sealed. Now, their capacitances are C'$_1$=(102.2±0.1)pF and C'$_2$ =(53.6±0.1)pF respectively at room temperature. We put the two capacitors into a temperature-control stove, raise the temperature of the stove very slow and keep at T$_1$ =(473±0.5)K for 3 hours. It means that the readings of capacitance are obtained under the condition of Cs or Hg saturated vapor pressure. Two comparable experimental curves are shown in Fig.2. From Ref.[15], the saturated pressure of Hg vapor P$_{Hg}$ =2304.4 Pa at 473K and measuring capacitance C'$_{2t}$= (56.4± 0.1) pF. From the ideal gas law, the density of Hg vapor is N$_{Hg}$ = P$_{Hg}$ /kT$_1$=3.53×10$^{23}$ m$^{-3}$. The formula of saturated vapor pressure of Cs vapor is **P=10$^{6.949-3833.7/T}$** psi and the effective range is 473K ≤T ≤623K[15]. We obtain the saturated pressure of Cs vapor P$_{Cs}$=0.0698 psi=481.3Pa at 473K and measuring capacitance C'$_{1t}$ = (3944±10) pF. The density of Cs atoms is N$_{Cs}$ = N$_1$ = P$_{Cs}$ /kT$_1$ =7.37×10$^{22}$ m$^{-3}$. The experiment shows that the number density N$_{Hg}$ of Hg vapor is 4.79 times as N$_{Cs}$ of Cs vapor but the capacitance change(C'$_{2t}$ − C'$_{20}$) of Hg vapor being only 1/1080 of (C'$_{1t}$ − C'$_{10}$) of Cs vapor! So unlike Cs atom, a Hg atom is non-polar atom.

The second experiment: investigation of the relationship between the electric susceptibility( x$_e$) of Cs or Hg vapor and temperatures(T) under each fixed density. The experimental apparatus was a closed glass container containing Cs vapor. Two stainless steel tubes **a** and **b** build up the glass cylindrical capacitor (Fig.3). Since R-r=H<<r, the capacitor could be approximately regarded as a parallel-plane capacitor. The capacitance was still measured by the digital meter, and vacuum capacitance C$_0$=(66.0±0.1)pF(where S$_2$ = (5.7±0.1)×10$^{-2}$ m$^2$, H$_2$=(7.6±0.1)mm ). Then the capacitor filled with Cs vapor at the fixed density N$_2$ and was put into the stove. By measuring the electric susceptibility x$_e$ of Cs vapor at different temperature T, we obtain x$_e$ =A+B/T≈B/T, where the slope B=(320±4) K and A≈0 . The capacitance of an identical capacitor containing Hg vapor was measured and we found that the slope is nearly zero, B≈0.0K( x$_e$ <0.07). The experimental results are shown in Fig.4.

The third experiment: measuring the capacitance of Cs vapor at various voltages (V) under a fixed N$_2$ and T$_2$. The apparatus was the same as the second experiment (C$_0$=(66.0±0.1)pF) and the method is shown in Fig.5. C was the capacitor filled with Cs vapor to be measured, which was kept at T$_2$= (323±0.5)K, and C$_d$ was used as a standard capacitor. Two signals V$_c$(t)=V$_{co}$ cosωt and V$_s$(t)=V$_{so}$ cosωt were measured by a digital oscilloscope having two lines. The two signals had the same frequency and always the same phase at different voltages when the frequency is higher than a certain value. It indicates that capacitor C filled with Cs vapor was the same as C$_d$ , a pure capacitor without loss. From Fig.5, we have C = C$_d$ (V$_{so}$ −V$_{co}$)/V$_{co}$. In the experiment V$_{so}$ can be adjusted from zero to 800V. The capacitance C at various voltages was shown in Fig.6. When V$_{co}$=V$_1$=(0.3±0.002)volt, C$_1$ =(130.0±0.2)pF is approximately constant, and x$_e$=C$_1$/C$_o$ − 1=0.9697. With the increase of voltage, the capacitance decreases gradually. When V$_{co}$=V$_2$=(560±2)V, C$_2$ =(68.0±0.2) pF, it approaches saturation and x$_e$ =C$_2$ /C$_0$ − 1≤0.0303≈0. If nearly all dipoles in a gas turn toward the direction of the field, this effect is called the saturation polarization. The C-V curve shows that the saturation polarization of Cs vapor is obvious when E= V$_2$ /H$_2$ ≥7.4×10$^4$v/m.

**C. Polarization theory of polar atoms** The electric susceptibility of gaseous polar molecules is[7]

$$x_e = N α + N d_o L(a)/ ε_o E \qquad (10)$$

Next we will consider how this equation is applied to Cs atoms. The induced dipole moment of an Cs atom is d$_{int}$ =G ε$_o$ E, where G=59.6×10$^{-30}$m$^3$ [21]. Due to the most field intensity is E≤7.4×10$^4$v/m in the experiment, then d$_{int}$ ≤3.9×10$^{-35}$ C.m =2.4×10$^{-14}$e.cm can be neglected. From Eq.(10) we obtain

$$x_e = N d L(a)/ ε_o E \qquad (11)$$

where d is the EDM of an Cs atom and N is the number density of Cs vapor. **L(a)= < cos θ > is the percentage of polar atoms which lined up with the field in the total number .** Note that x$_e$ = ε$_r$ − 1= C/C$_o$ − 1, where ε$_r$ is the dielectric constant , E=V/H and ε$_o$= C$_o$ H / S, leading to **the polarization equation of Cs atoms**

$$C − C_o = β L(a)/a, \qquad (12)$$

where β = S N d$^2$/kTH is a constant. Due to a=d E/k T= dV/kTH **we obtain the first formula of atomic EDM**

$$d_{atom} = (C - C_o)V / L(a)SN \tag{13}$$

In order to work out **L(a)** and **a** of the first experiment, note that in the third experiment when the field is weak ($V_1$=0.3volt), $a_1 \ll 1$ and $L(a_1) \approx a_1/3$, $C_1 - C_o = \beta/3$ and $\beta = 192$ pF. When the field is strong($V_2$= 560volt), $a_2 \gg 1$ and $L(a_2) \approx 1$, $C_2 - C_o = L(a_2)\beta/a_2 \approx \beta/a_2$. We work out $a_2 \approx \beta/(C_2 - C_o) = 96$, $L(a_2) \approx L(96) = 0.9896$, $a_2 = \beta L(a_2)/(C_2 - C_o) = 95$, $L(a_2) = L(95) = 0.9895$. Due to **a**=d E/kT=dV/kTH, so $a/a_2 = VT_2H_2/T_1H_1V_2$. Substituting the corresponding values, we obtain **a**=0.09171 and L(**a**)≈**a**/3=0.0306. L(**a**)≈0.0306 means that only 3.06% of Cs atoms have be oriented in the direction of the field when the field $E=V/H_1=1.1\times 10^2$V/m. Substituting the values: $S_1=5.8\times 10^{-2}$ m$^2$, $N_1=7.37\times 10^{22}$ m$^{-3}$, V=1.0volt, $C - C_o = C'_{1t} - C'_{10} = 3890$pF and L(a), we work out

$$d(Cs) = (C - C_o)V / L(a) S_1 N_1 = 2.97\times 10^{-29} \text{C.m} = 1.86\times 10^{-8} \text{e.cm} \approx 3.5\, e\, a_o \tag{14}$$

The statistical errors is $\triangle d_1/d = \triangle C/C + \triangle C_o/C_o + \triangle S_1/S_1 + \triangle V/V + \triangle N_1/N_1 < 0.12$, considering all systematic error $\triangle d_2/d < 0.08$ (including $\triangle L(a)/L(a)$), and the combination error $\triangle d/d < 0.15$. We find that

$$\mathbf{d(Cs) = [2.97\pm 0.36(stat) \pm 0.24(syst)]\times 10^{-29} \text{C.m} = [1.86\pm 0.22(stat)\pm 0.15(syst)]\times 10^{-8}\text{ e. cm}} \tag{15}$$

**Although above calculation is simple, but no physicist completed the calculation up to now!**

## 4. Discussion

① Some evidence for CP violation beyond the Standard Model comes from Cosmology. Astronomical observations indicate that our Universe is mostly made of matter and contains almost no anti-matter. The first example of CP violation was discovered in 1964, but it has been observed only in the decays of the $K_o$ mesons. After 38 years, the BaBar experiment at Stanford Linear Accelerator Center (SLAC) and the Belle collaboration at the KEK laboratory in Japan announced the second example of CP violation. "The results gave clear evidence for CP violation in B mesons. However, the degree of CP violation now confirmed is not enough on its own to account for the matter-antimatter imbalance in the Universe." "In the past few years we have learned to cleanly isolate one event in a million." (SLAC Press Release July 23, 2002 and Sep 28, 2006). Theorists found it hard to see why CP symmetry should be broken at all and even harder to understand why any imperfection should be so small. This fact suggests that there must be other ways in which CP symmetry breaks down and more subtle effects must be examined. So EDM experiments are now considered an ideal probe for evidence of new sources of CP violation. If an EDM is found, it will be compelling evidence for the existence of new sources of CP violation. **Our experiments showed that new example of CP violation occurred in K, Rb or Cs atoms and it is a classic example of how understanding of our universe advances through atomic physics research.**

② The formula $d_{atom} = (C - C_o)V/L(a)SN$ can be justified easily. The magnitude of the dipole moment of an atom is d = e r. N is the number of atoms per unit volume. L(a) is the percentage of Cs atoms lined up with the field in the total number. Suppose that the plates of the capacitor have an area S and separated by a distance H, the volume of the capacitor is SH. When a field is applied, the Cs atoms tend to orient in the direction of the field. On the one hand, the change of the charge of the capacitor is $\triangle Q=(C-C_0)V$. On the other hand, when the Cs atoms are polarized by the orientation, the total number of Cs atoms lined up with the field is SHNL(a). The number of layers of Cs atoms which lined up with the filed is H/r. Because the positive and negative charges cancel out each other inside the Cs vapor, the polarization only gives rise to a net positive charge on one side of the capacitor and a net negative charge on the opposite side. Then the change $\triangle Q$ of the charge of the capacitor is $\triangle Q = SHN L(a)e/(H/r) = SN L(a)d = (C - C_0)V$, so the EDM of an Cs atom is $\mathbf{d = (C - C_0)V/ SN L(a)}$.

③ If Cs atom is the polar atom, we should calculate the EDM of an Cs atom using the same method such as water molecules from $d = (3k \varepsilon_o B / N_2)^{1/2}$. Note that $a_1 = a_2 V_1/V_2 = 0.0509$, the polarization is respectively

$$P_1 = \varepsilon_o \chi_e E_1 = (C_1 - C_o)V_1/S_2 = N_2 d L(a_1) = N_2 d a_1/3 = 0.017 N_2 d \approx N_2 d/58.9 \tag{16}$$

$$P_2 = \varepsilon_o \chi_e E_2 = (C_2 - C_o)V_2/S_2 = N_2 d L(a_2) = 0.9895 N_2 d \approx N_2 d \tag{17}$$

Due to the density $N_2$ is unknown in the second and the third experiments, from Eq.(13) and Eq.(17) we can obtain

$$N_2 = (C_2 - C_o)V_2 L(a) S_1 N_1 / (C - C_o)V L(a_2) S_2 = 6.70\times 10^{20} \text{m}^{-3} \tag{18}$$

From $d = (3k\varepsilon_o B/N_2)^{1/2}$, note that $k = dE/aT_1 = dV/aT_1H_1$, **we obtain the second formula of atomic EDM**

$$\mathbf{d_{atom} = (3k \varepsilon_o B / N_2)^{1/2} = 3 V \varepsilon_o B / aT_1H_1N_2} \tag{19}$$

Substituting the values: B=320K, $N_2$ =6.70×$10^{20}$ $m^{-3}$, V=1.0volt, $T_1$=473K, $H_1$=9.6mm and a=0.09171, we work out

$$d(Cs)=3.04×10^{-29} C.m= 1.90×10^{-8} e.cm \tag{20}$$

Using two different methods and different experimental data we obtain the same result, it proved that the data are reliable and the EDM of an Cs atom has been measured accurately.

④ When the saturation polarization appeared, due to the angle θ = 0, then <cos θ >= L(a)≈1 and this will occur only if a>>1[7]. Because the most EDM of polar molecules $d_o$ is roughly $10^{-29}$ C.m and the breakdown field intensity of gaseous dielectric is roughly $10^7$v/m. The most potential energy $d_oE$ of a molecule is about 6.3 ×$10^{-4}$ ev. The average kinetic energy KT of each molecule in a gas is about 0.03ev at ordinary temperatures (T=300k), and most value $a_{max}=d_oE/KT$ <0.03<<1. It means that the saturation polarization can happen only if temperatures near absolute zero for all polar molecules. So no scientist has observed the saturation polarization of any gaseous dielectric till now. R.P. Feynman stated that *" when a filed is applied, if all the dipoles in a gas were to line up, there would be a very large polarization, but that does not happen"* [8]. The saturation polarization of K, Rb or Cs vapor in ordinary temperatures is an entirely unexpected discovery[16-18].

⑤ In Eq.(12) let the new function

$$f(a)= L(a)/a= [(e^a + e^{-a}) / a ( e^a – e^{-a})] – 1/a^2 \tag{21}$$

From f ″(a)=0, we work out the knee of the function L(a)/a at

$$a=1.9296813≈1.93 \tag{22}$$

Corresponding knee voltages $V_k$=11.4V and the knee field $E_k=V_k/H_2$≈1.5×$10^3$V/m. By contrast with the curve in Fig.6, it is clear that our polarization equation is valid. When the saturation polarization of Cs vapor occurs, nearly all Cs atoms (more than 98.9%) turned toward the direction of the field, and C≈$C_o$ or **its dielectric property is nearly the same as vacuum.** On the other hand, when all Cs atoms are lined up with the field, Cs vapor result in the largest polarization Nd and the energy density $U_m$ of the system is given by $U_m$ = - NdE.. Cs vapor just exist in the lowest energy state. In quantum field theory, vacuum as the source of asymmetry, it is defined as the lowest energy state of a system, so we see the vacuum state containing a large number of Cs atoms! Table 2 provided a complete analysis of the saturation polarization of Cs atoms at different voltages.

**Table 2 The orientation polarization of ground state Cs atoms at different voltages(V)**

| Volt | 0.3v | 50v | 100v | 560v | 2000v | 4000v | ∞ |
|---|---|---|---|---|---|---|---|
| E(V/m) | 39.5 | 6.6×$10^3$ | 1.3×$10^4$ | 7.4×$10^4$ | 3.1×$10^5$ | 6.2×$10^5$ | ∞ |
| a | 0.051 | 8.48 | 17.0 | 95 | 339 | 679 | ∞ |
| L(a) | 0.017 | 0.882 | 0.941 | 0.9895 | 0.997 | 0.9985 | 1.0 |
| $x_e$ | 0.9697 | 0.3026 | 0.161 | 0.0303 | 0.0086 | 0.0043 | 0.0000 |
| $ε_r$ | 1.9697 | 1.3026 | 1.161 | 1.0303 | 1.0086 | 1.0043 | 1.0000 |
| $C_o$(pF) | 66.0 | 66.0 | 66.0 | 66.0 | 66.0 | 66.0 | 66.0 |
| C'(pF) | 130.0 | 85.97 | 76.6 | 68.0 | 66.56 | 66.28 | 66.0 |
| P | 0.017Nd | 0.882Nd | 0.941Nd | 0.9895Nd | 0.997Nd | 0.9985Nd | Nd |
| $U_m$ | -0.017NdE | -0.882NdE | -0.941NdE | -0.9895NdE | -0.997NdE | -0.9985NdE | -NdE |

where L(a) = [($e^a + e^{-a}$) / ( $e^a – e^{-a}$)] – 1/a, the polarization P =Nd L(a), the energy density $U_m$ = -Nd EL(a), N= 6.70×$10^{20}$$m^{-3}$, T= 323K, a= $a_2$ V/ $V_2$, $V_2$=560v, $a_2$=95, E=V/H, H=7.6mm, β =192 pF.

⑥ The saturation polarization of K, Rb or Cs vapor can not be explained by existing theories and it tells us that the mechanism of polar atoms by which orientation polarization arises completely differs from polar molecules. The orientation polarization of the polar molecule, such as HCl or $H_2O$ etc, is the molecule as a whole turned toward the direction of an external field. We can measure the average rotation energy of individual molecule by using kT. So we can work out the probability distribution at thermal equilibrium using the Boltzmann factor exp (–U/kT). Unlike polar molecules, the orientation polarization of the polar atom, such as K, Rb or Cs atom, is only the valence electron in the outermost shell turned toward the direction of the field,



but the rest of the atom(include the nucleus) does not move ! Due to the rotational inertia of an electron is much less than an atom, when finding the orientation probability distribution of polar atoms at thermal equilibrium, we should use the new Boltzmann constant $k$ instead of k=1.38×$10^{-23}$ J/K. Its form is

$$k = dE/aT \tag{23}$$

We will measure the average rotation energy of polar atoms by using dE/a. Table 3 gives three values of new Boltzmann constant for Cs atom, where d= 2.97×$10^{-29}$ C.m.

**Table 3 The three values of new express of Boltzmann constant for Cs atom**

| T(K) | V | H | a | E | k |
|---|---|---|---|---|---|
| 323 | 0.3 | 7.6 mm | 0.0509 | 39.5 V/m | 7.13×$10^{-29}$ J/K |
| 323 | 560 | 7.6 mm | 95 | 7.37×$10^{4}$ V/m | 7.13×$10^{-29}$ J/K |
| 473 | 1.0 | 9.6 mm | 0.09171 | 104.2 V/m | 7.13×$10^{-29}$ J/K |

Due to k=7.13×$10^{-29}$ J/K<< 1.38×$10^{-23}$ J/K, so the saturation polarization of polar atoms is easily observed.

⑦We will calculate new value of Boltzmann constant for water molecule. From $k$= dE/aT and $x_e$'= $x_e$ –A= $Nd_o^2/3k\varepsilon_oT$, We obtain k= $(P/3x_e'\varepsilon_o)^{1/2}$ d/T. Table 4 gives four values of new Boltzmann constant for water.

**Table 4 The four values of new express of Boltzmann constant for water molecule**

| T(K) | Pressure(cm Hg) | $x_e$' | k |
|---|---|---|---|
| 393 | 56.49 | 0.003802 | 1.380×$10^{-23}$ J/K |
| 423 | 60.93 | 0.003517 | 1.385×$10^{-23}$ J/K |
| 453 | 65.34 | 0.003288 | 1.385×$10^{-23}$ J/K |
| 483 | 69.75 | 0.003087 | 1.385×$10^{-23}$ J/K |

where d= 6.28×$10^{-30}$ C.m. Clearly, when the new form of Boltzmann constant was applied to any molecule, it will gained the same results as usual statistical mechanics. **Its success proved that the true significance of Boltzmann constant is $k$= dE/aT rather than k=1.38×$10^{-23}$ J/K.**

⑧The shift in the energy levels of an atom in an electric field is known as the Stark effect. Normally the effect is quadratic in the field strength, but first excited state of the hydrogen atom exhibits an effect that is linear in the strength. This is due to the degeneracy of the excited state. This result shows that the hydrogen atom (the quantum number n=2 ) is polar atom with large EDM, $d_H$=3e$a_o$=1.59×$10^{-8}$e.cm. This EDM does not depend on the field strength $\varepsilon$, hence it is not induce by the external field but is inherent behavior of the atom[13,14]. According to the Boltzmann distribution, the number of particles with energies in the interval $E$～$E+dE$ in thermal equilibrium is given by

$$ndE = NZ(T)\exp(-U/kT)g(E)dE \tag{24}$$

where n is the number density of particles, N is the total number of particles, Z(T) is the partition function, g($E$) is the density of states. For example, in the sun, with a surface temperature of 6000K, only $10^{-8}$ of the hydrogen atoms in the in the solar atmosphere are in the n=2 state.

The electron of hydrogen atoms must be previously raised to the higher state by an input of energy. From 3kT/2 ≥ (-3.4eV)-(-13.6eV)=10.2eV, we find the temperature of the atmosphere T≥7.9×$10^{4}$K. The temperature is so high, that nearly all hydrogen atoms are in the first excited state. Due to the saturation polarization of polar atoms, such as first excited state of the hydrogen atoms, is easily observed, so a vacuum state containing a large number of hydrogen atom of excited state may occur in early Universe.

Although the experiments are simple, but no physicist completed the experiments up to now! Our measuring process and experimental results can be easily repeated in any laboratory because of the details of the experiments are described in the article. Our experimental apparatus are still kept, we welcome anyone who is interested in the experiments to visit and examine it.

**Acnowledgement**   The authors thank to our colleagues Ri-Zhang Hu, Zhao Tang , Rui-Hua Zhou, Shao-Wei Peng, Yong Chen, Xue-ming Yi, Xiao-ming Wang, Xing Huang, and Engineer Jia You for their help in the work.

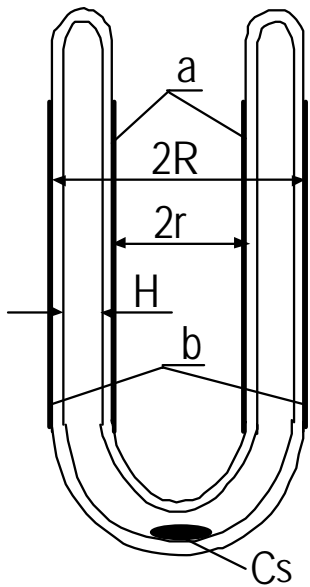
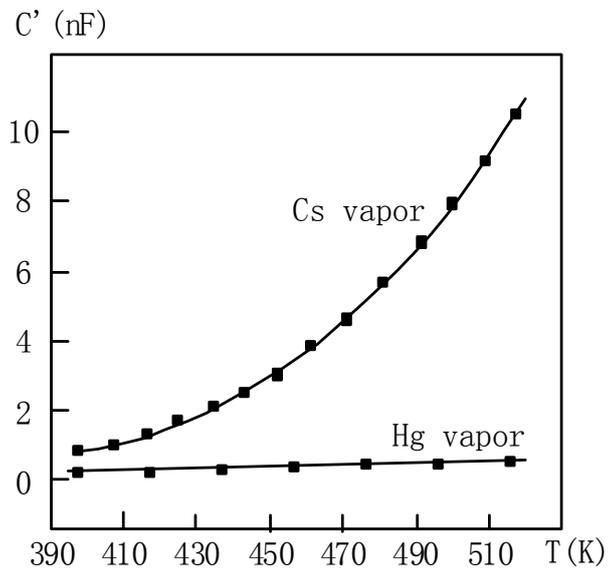

**Fig.1** This is the longitudinal section of the first experimental apparatus. Silver layers **a** and **b** build up a cylindrical glass capacitor. (not to scale).

**Fig.2** Two curves showed that the relationship between the capacitance C' of Cs or Hg vapor and the density N respectively, where 1nF=$10^3$ pF.

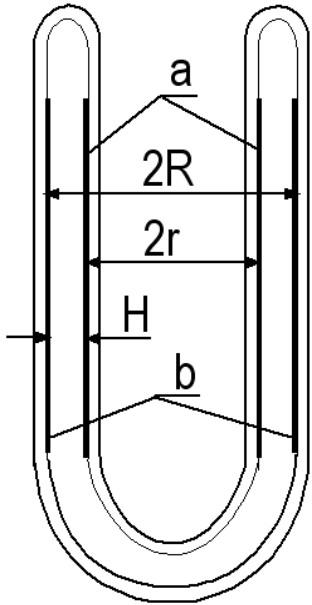

**Fig.3** This is the longitudinal section of another two experimental apparatus. Two round stainless steel tubes **a** and **b** build up a glass capacitor (not to scale).

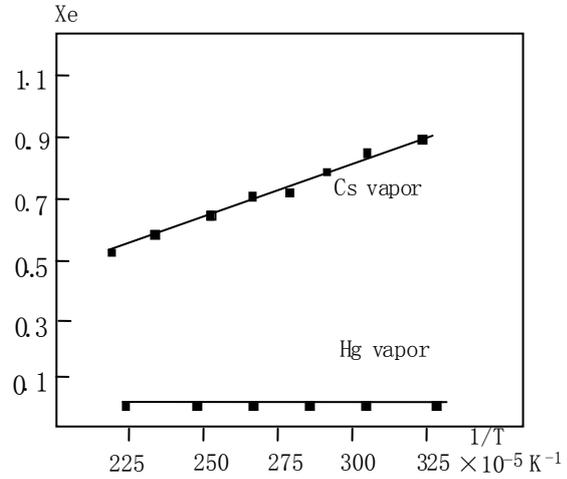

**Fig.4** The temperature (T) dependence of the susceptibility ($x_e$) of Cs or Hg vapor. The slope of Cs vapor is B≈320(k) but the slope of Hg vapor is nearly zero, B≈0.0(k).

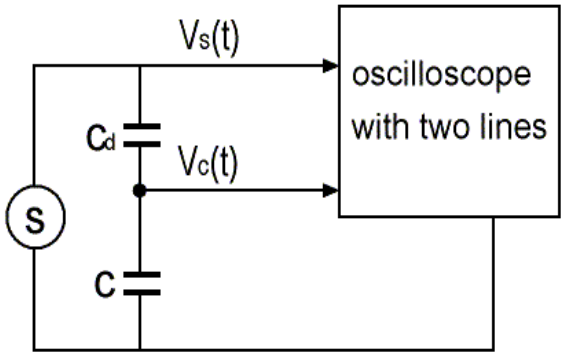

**Fig.5** The diagram shows the experimental method, in which C is capacitor filled with Cs vapor to be measured and Cd is a standard one, where $V_s(t) = V_{so} \cos \omega t$ and $V_c(t) = V_{co} \cos \omega t$.

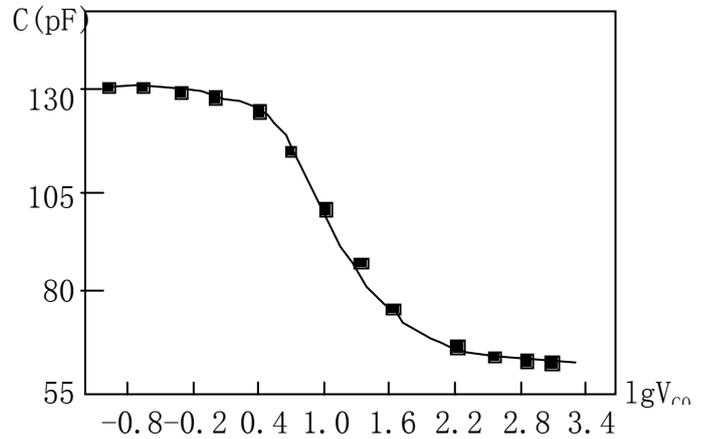

**Fig.6** The experimental curve shows that the saturation polarization effect of the Cs vapor is obvious when $E \geq 7.4 \times 10^4$ v/m.